\documentclass[12pt,nofootinbib,aps,prd]{revtex4-2}
\usepackage[T1]{fontenc}
\usepackage{graphicx}
\usepackage{multirow}
\usepackage{amssymb}
\usepackage[usenames]{color}
\usepackage[toc,page]{appendix}
\usepackage{slashed}
\usepackage{bbding}
\usepackage{booktabs}
\usepackage{float}
\usepackage{colordvi}
\usepackage{amsmath}
\usepackage{amssymb}
\usepackage{amsfonts}
\usepackage[dvipsnames]{xcolor}
\usepackage[colorlinks=true,urlcolor=blue,linkcolor=blue,citecolor=blue]{hyperref}
\usepackage{cancel}
\usepackage{ulem}
\usepackage[top=1.135in,bottom=1.17in,left=1.16in,right=1.16in]{geometry}
\usepackage{silence}
\numberwithin{equation}{section}
\allowdisplaybreaks

\newcommand{\eq}[1]{\begin{align}#1\end{align}}

\usepackage[dvipsnames]{xcolor}

\begin{document}
\thispagestyle{empty}

\title{Authentic Majorana versus singlet Dirac neutrino contributions to $\mu^{+}\mu^{+}\to \ell^{+}\ell^{+}$  $(\ell=e,\tau)$ transitions}

\vspace{50pt}

\author{Jorge Luis Gutiérrez Santiago}\email{jorge.gutierrez@cinvestav.mx}\affiliation{Departamento de F\'isica, Centro de Investigaci\'on y de Estudios Avanzados del Instituto Polit\'ecnico Nacional\\
Apartado Postal 14-740, 07360 Ciudad de M\'exico, M\'exico}\author{G. Hern\'andez-Tom\'e}\email{gerardo_hernandez@uaeh.edu.mx}\affiliation{Área Académica de Matemáticas y Física, Universidad Autónoma del Estado de Hidalgo,
Carretera Pachuca-Tulancingo Km. 4.5, C.P. 42184, Pachuca, Hidalgo, México.}
\author{Diego Portillo-Sánchez}\email{diego.portillo@cinvestav.mx}\affiliation{Departamento de F\'isica, Centro de Investigaci\'on y de Estudios Avanzados del Instituto Polit\'ecnico Nacional\\
Apartado Postal 14-740, 07360 Ciudad de M\'exico, M\'exico}
    \author{Javier Rendón}
\email{jesus.rendon@correo.nucleares.unam.mx}
\affiliation{Instituto de Ciencias Nucleares,
Universidad Nacional Aut\'onoma de M\'exico,
A.P. 70-543, Mexico City 04510.}

\begin{abstract}
We revisited the computation of the cross section for the $\mu^+\mu^+\to \ell^+\ell^{+}$ $(\ell^+=\tau^+,\, e^+)$ transitions in the presence of new heavy Majorana neutrinos. We focus on scenarios where the masses of the new states are around some few TeV, the so-called low-scale seesaw models. Our analysis is set in a simple model with only two new heavy Majorana neutrinos considering the current limits on the heavy-light mixings. Our results illustrate the difference between the genuine effects of two nondegenerate heavy Majorana states and the degenerate case where they form a Dirac singlet field.
\end{abstract}

\maketitle

\section{Introduction}

The observation of neutrino oscillation \cite{Super-Kamiokande:1998kpq, SNO:2001kpb, SNO:2002tuh}  establishes that lepton flavor is not conserved, at least in the neutrino sector, and demands an explanation of the origin and nature of neutrino masses \footnote{Testing whether neutrinos are Dirac or Majorana particles remains one of the most significant challenges in particle physics. The most promising way is the search for double beta decay of nuclei without neutrinos \cite{Dolinski:2019nrj, Rodejohann:2011mu, Vergados:2012xy}. If this process were observed, it would provide a direct evidence that neutrinos are Majorana particles. Furtheremore, alternative methods have been discussed in the literature, see for example \cite{Rodejohann:2017vup, Marquez:2024tjl}.}. Furthermore, if neutrinos are massive particles, they will induce charged lepton flavor violation (cLFV) at the one-loop level, even though these processes have never been observed. This fact can be well explained in minimal massive neutrino models such as both the $\nu$SM \cite{Mohapatra:1998rq} and type-I seesaw models  \cite{Minkowski:1977sc, Yanagida:1979as, Gell-Mann:1979vob, Mohapatra:1979ia}. In the former, there is a suppression factor of $(m_\nu/E)^2$ (with $E$ as the scale of the processes) for the contribution of the light active neutrinos, while in the latter, the contributions of the new heavy sector have a suppression of the order $(E/m_N)^2$ in the heavy-light mixing of the heavy states, which is equally small.

In contrast, cLFV can emerge considerably in other well-motivated massive neutrino scenarios known as the low-scale seesaw models. Specifically, the inverse \cite{Mohapatra:1986bd, Gonzalez-Garcia:1988okv} and linear models \cite{Malinsky:2005bi} can be seen as an alternative to the traditional high-scale Type-I seesaw by considering specific textures for the neutrino mass matrix \cite{CentellesChulia:2024uzv}. In this way, the search for cLFV processes is one of the most compelling ways to test such scenarios \cite{Garnica:2023ccx, Forero:2011pc, Abada:2023raf, Celestino-Ramirez:2024gmq}.

Nowadays, the so-called three ``golden'' cLFV processes $\mu\to e\gamma$, $\mu\to e\bar{e}e$, and $\mu-e$ conversion set limits with sensitivities around of the order $10^{-13}$ for the former two and $10^{-12}$ for the latter process \cite{MEG:2013oxv, SINDRUM:1987nra, SINDRUMII:2006dvw}, respectively. Nevertheless, a completely new generation of experiments would be able to considerably improve such limits. Specifically, the MEG-II, Mu3e, PRISM, COMET, and Mu2e collaborations will reach sensitivities of $6\times10^{-14}$ \cite{MEGII:2018kmf, Cavoto} for $\mu\to e\gamma$, $10^{-16}$ \cite{Blondel:2013ia} for $\mu\to e\bar{e}e$, $10^{-18}$ \cite{Alekou:2013eta} for $\mu-e$(Ti) conversion, and $10^{-17}$ \cite{Kuno:2013mha} and $10^{-16}$ \cite{Mu2e:2014fns} for $\mu-e$(Al) conversion, respectively \footnote{Searches for cLFV in other sectors such as tau \cite{BaBar:2009hkt, Hayasaka:2010np, Belle:2007cio, BaBar:2006jhm, Belle:2011ogy, ParticleDataGroup:2022pth}, mesons  \cite{KTeV:2007cvy, BNL:1998apv, Sher:2005sp, BESIII:2013jau, BES:2004jiw, BaBar:2006tnv, BaBar:2012azg, LHCb:2013tno, BaBar:2008pet, CLEO:2008lxu}, $Z$  \cite{ATLAS:2014vur, OPAL:1995grn, DELPHI:1996iox}, and Higgs decays \cite{CMS:2016cvq, CMS:2017con} play also a crucial role in the intense activity of both experimental and theoretical groups. These transitions are complementary for exploring effects at different energy scales and may help to distinguish different sources of cLFV among the plethora of new physics scenarios.}.

Additionally, the potential advent of muon colliders \cite{Gallardo:1996aa, Geer:2009zz, Delahaye:2019omf, MICE:2019jkl, EuropeanStrategyforParticlePhysicsPreparatoryGroup:2019qin, Long:2020wfp,  AlAli:2021let, ANTONELLI2016101, Accettura:2023ked} opens another route for complementary searches of new physics (NP) \cite{Rodejohann:2010jh, Rodejohann:2010bv, NeutrinoFactory:2002azy,  Shiltsev:2010qg, Buttazzo:2018qqp, Chiesa:2020awd}. This idea includes the possibility of both muon-antimuon ($\mu^- \mu^+$) \cite{Delahaye:2019omf, MICE:2019jkl, EuropeanStrategyforParticlePhysicsPreparatoryGroup:2019qin, Long:2020wfp, Accettura:2023ked} and same-sign ($\mu^+\mu^+$) muon colliders \cite{Heusch:1995yw, Hamada:2022mua, Hamada:2022uyn, Fridell:2023gjx, Dev:2023nha} with a center-of-mass energy in the TeV scale. Motivated by this, in this work, we compute the cross section for the $\mu^+\mu^+\to \ell^+\ell^{+}$ ($\ell^+=\tau^+,\, e^+$) transitions in scenarios that introduce new heavy neutrino states. Our attention is focused on models with massive Majorana neutrinos around the TeV scale. Furthermore, our results address some comments on previous computations for the $e^-e^-\to \mu^-\mu^-$ \cite{Cannoni:2002ny} and $\mu^\pm\mu^\pm\to \ell^\pm\ell^\pm$ ($\ell^\pm=\tau^\pm,\, e^\pm$) channels \cite{Yang:2023ojm}. In those previous works, only the contributions of one-loop box diagrams with explicit lepton number-violating (LNV) vertices are included. However, we highlight that an accurate estimation requires another set of diagrams omitted in Refs. \cite{Cannoni:2002ny, Yang:2023ojm} \footnote{Note that our computation can be implemented easily to other specific low-scale-seesaw models such as the linear and the inverse seesaw models.
}. 

The manuscript is structured as follows: Section \ref{sec: the model} introduces a simple model that captures all the relevant effects of new heavy neutrinos to cLFV transitions. This minimal scenario will allow us to estimate the genuine effects of two nondegenerate Majorana states and compare it with the degenerate case, where they define a Dirac singlet field. After that, 
 section \ref{sec: computation} is devoted to the meaningful aspect of our computation introducing the $\mu^+\mu^+\to\ell^+\ell^+$ ($\ell=e,\, \tau$) one-loop amplitudes in massive neutrino models (some important details and identities used in our computation are left to the appendixes). Then section \ref{NE} estimates the total cross section for these processes based on both the current limits of the heavy-light mixings and a perturbative condition. Finally, the conclusions and a summary of our results are presented in section \ref{Conclusions}.

\section{The model}\label{sec: the model}

We work in a simplified model previously presented in Refs. \cite{Ilakovac:1994kj, Hernandez-Tome:2019lkb, Hernandez-Tome:2020lmh}. Here, the neutrino sector is composed of five self-conjugate states whose left-handed components $\chi_{Li}$ include the three active neutrinos ($i = 1, 2, 3$) plus two sterile spinors $\chi_{\{4,5\}}$ of opposite lepton number. The Lagrangian defining the neutrino masses is assumed to have the form:
\eq{
-\mathcal{L}\supset\,\sum_i^3 Y_i \tilde{\Phi}^{\dagger} L_i \overline{\chi_5^c}+M\, \overline{\chi_5^c}\chi_{4} +\frac{1}{2}\mu \, \overline{\chi}_5 \chi_5^c+ \textrm{H.c.} \, , \label{eq: MassLagrangian} 
} where $\tilde{\Phi}\equiv i\sigma_2\Phi^{*}$ stands for the charge conjugate Higgs field, $L_i$ the lepton SM doublets, and $Y_i$ corresponds to Yukawa couplings. After the spontaneous electroweak symmetry breaking, the neutrino mass matrix is written as
\eq{
\mathcal{M^{\nu}}=\left(\begin{tabular}{ccccc}
     $0$ & $0$ & $0$ & $0$ & $m_1$ \\
     $0$ & $0$ & $0$ & $0$ & $m_2$ \\
     $0$ & $0$ & $0$ & $0$ & $m_3$ \\
     $0$ & $0$ & $0$ & $0$ & $M$ \\
     $m_1$ & $m_2$ & $m_3$ & $M$ & $\mu$ \\
\end{tabular}\right)\label{eq: massmatrix}, 
 }with the masses $m_i\equiv Y_i v/\sqrt{2}$ for ($i=1,2,3$), and $v\approx256\,$GeV for the vacuum expectation value (VEV) of the Higgs field.  The diagonalization of Eq. (\ref{eq: massmatrix}) leads to three massless neutrinos $\nu_i$ and two heavy neutrino states with masses
 \eq{
 m_{\{N_1,N_2\}}&=\frac{1}{2}\left(\sqrt{4(m^2+M^2)+\mu^2}\pm \mu\right)\, \label{eq: Heavy masses},}
with the definition $m^2\equiv m_1^2+m_2^2+m_3^2$.

On the other hand, the weak charged lepton currents necessary for computing the cLFV processes are described by \cite{Ilakovac:1994kj, Hernandez-Tome:2019lkb}  (arising, at one-loop level, through the mixing of the leptonic sector)  
\eq{
\mathcal{L}_{W^{\pm}}=-\frac{g}{\sqrt{2}}W_\mu^-\sum_{i=1}^3\sum_{j=1}^5 B_{ij}\,\bar{\ell}_i\gamma_\mu P_L \chi_j + \textrm{H.c.}\,.\label{eq: Charged-current Lagrangian}
}Working in the Feynman-'t Hooft gauge would also require considering the interaction of the unphysical charged Goldstone bosons with a pair of leptons, those vertices are described by  \cite{Ilakovac:1994kj, Hernandez-Tome:2019lkb}
\eq{
\mathcal{L}_{G^{\pm}}=-\frac{g}{\sqrt{2}}G^-\sum_{i=1}^3\sum_{j=1}^5 B_{ij}\,\bar{\ell}_i\left(\frac{m_{\ell_i}}{M_W} P_L-\frac{m_{X_j}}{M_W}P_R\right) \chi_j + \textrm{H.c.}\,.
\label{eq: Goldstone Lagrangian}}
In the above expressions, $B_{ij}$ is a  rectangular $3\times 5$ matrix defining the mixing in the leptonic sector. The elements involving the heavy new states $N_{1,2}$ are crucial in the description of both lepton number violating (LNV) and cLFV effects, they can be written as:
\eq{
B_{kN_1}=-i\frac{r^{1/4}}{\sqrt{1+r^{1/2}}}s_{\nu_k},\quad B_{kN_2}=\frac{1}{\sqrt{1+r^{1/2}}}s_{\nu_k},
}with the definition $r\equiv m_{N_2}^2/m_{N_1}^2$, and $s_{\nu_k}$ ($k=e,\, \mu,\, \tau$) defining the angles of the heavy states with the three flavors, they are expressed as follows
\eq{
s_{\nu_k}^2=\frac{m_k}{\sqrt{m_{N_1}m_{N_2}}}.\label{eq: Heavy-Light mixings definition}
} 
Something appealing about this scenario is that all the phenomenology of the leptonic mixing is described in terms of only five independent parameters, namely; a heavy mass $m_{N_1}$, the ratio $r$ and the three heavy-light mixings $s_{\nu_k}$.

It is important to mention that this is a simplified model unable to explain the masses and mixings of the light active neutrinos, however; it suffices to describe the main motivation of this work, namely, the study of cLFV effects in the presence of new heavy neutrino states with masses around $\mathcal{O}$(TeV). The generation of small masses and light-light mixings for the three active neutrinos would require the addition of extra singlets \footnote{ In the original formulations of both the inverse and linear low-scale scenarios [14-16], the neutrino mass matrix has dimension $9\times 9$, as these scenarios require the addition of $6$ new neutral singlets. This complexity makes it impractical to derive complete analytical expressions for the unitary matrix $U$ that diagonalizes the neutrino mass matrix and for the physical masses of the neutrinos, leading to approximate or numerical analyses that may turn out complicated to follow.}. This could be accommodated with the usual mechanisms (in $\nu$SM or Type I seesaw models, where cLFV effects are strongly suppressed) or by considering specific textures for the neutrino mass matrix based on approximated symmetries as required in the low-scale seesaw scenarios.  Nevertheless, the extra spinors and couplings needed to generate light neutrino masses will have no further effects on cLFV observables, see references \cite{Hernandez-Tome:2019lkb, Hernandez-Tome:2020lmh}  for details. 

We remark that this parametrization allows us to describe,  in a simple way, the effects of two heavy Majorana states and study the limit of almost degenerate masses ($r\to 1$) forming a pseudo-Dirac field. We address this point in detail in section \ref{NE}  by providing an estimation of the maximal total cross section (consistent with the current limits on the heavy-light mixings and a perturbative condition for the Yukawa couplings).

\section{Computation}\label{sec: computation}

\begin{figure}[!ht]
\begin{center}
\begin{tabular}{cc}
\includegraphics[scale=.75]{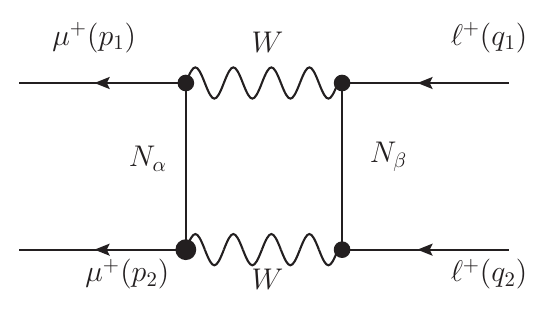}&
\includegraphics[scale=.75]{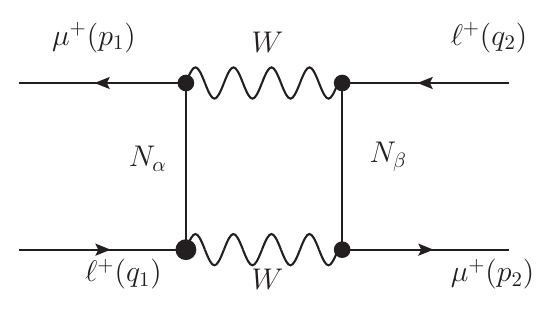} \\
{\small (a)} & {\small(b)} 
 \end{tabular}
 \caption{Feynman diagrams for the $\mu^+(p_1)\mu^+(p_2)\to \ell^+(q_1)\ell^+(q_2)$ $(\ell =e,\, \tau)$ transitions in the presence of heavy Majorana neutrinos. Working in the Feynman-'t Hooft gauge must include the four possible diagrams for the (a) and (b) contributions involving the $W$'s gauge bosons and the scalar-charged Goldstone vertices, see Appendix \ref{App: Amplitudes}.} \label{fig: Diagrams}
\end{center}
\end{figure}

In the presence of heavy Majorana neutrino states, Fig. \ref{fig: Diagrams} shows the two one-loop box diagrams contributing to the processes $\mu^+(p_1)\mu^+(p_2) \to \ell^+(q_1)\ell^{+}(q_2)$ ($\ell=e,\,\tau$). The diagrams (b) are present for an arbitrary vector-like lepton that mixes with the active neutrinos of the SM. Moreover, in type-I seesaw models, the existence of Majorana masses for the singlet fields for neutrinos inherently implies lepton number violation (LNV). These LNV effects give rise to the presence of box diagrams (a). Thus, we remark that a complete computation for these processes in type-I seesaw models (including their low-scale variants), must include both (a) and (b) contributions \footnote{Note that in the minimal $\nu$SM model, only contributions (b) are present, because in this model neutrinos are Dirac particles that generate their masses ($m_\nu\sim  \mathcal{O}$(eV)) in an analogous way to the rest of fermions, via Yukawa couplings with the Higgs field. However, in this scenario, the cLFV effects will be strongly suppressed by an analogue Glashow-Iliopoulos-Maiani-like mechanism.}.

As far as we know, previous works \cite{Cannoni:2002ny, Yang:2023ojm} have solely focused on evaluating the genuine LNV effects (diagrams (a)) considering effective couplings without discussing the reasons for omitting diagrams (b) or comparing (a) and (b) contributions.  In this work, we cover those points. Even more, as previously mentioned, in the type-I seesaw models only their low-scale versions (both linear and inverse) can potentially lead to observable cLFV rates. These scenarios are based on an approximated LNV symmetry, where the small LNV Majorana mass terms combined with larger Dirac masses lead to the formation of pseudo-Dirac neutrinos, namely, a pair of almost degenerate heavy neutral states, that in the degenerate limit form a Dirac singlet. We will study this limit considering the set-up presented in ref. \cite{Hernandez-Tome:2019lkb, Hernandez-Tome:2020lmh}.

By working in the approximation where the masses of the external charged leptons are neglected \footnote{Given that we will consider the collision energy in the regime of a few TeV, i.e., $m_{\ell}/\sqrt{s}\ll 1$.} and after some algebraic steps and the use of some Fierz identities (see Appendix \ref{App: Amplitudes} for further details), we have verified that the total amplitude $\mathcal{M}=\mathcal{M}_{(a)}+\mathcal{M}_{(b)}$ can be expressed in a very simple form as follows
\eq{
\mathcal{M}=\frac{\alpha_W}{16\pi}\frac{e^2}{s_W^2}  F(s,t)\,\Gamma_R\odot\Gamma^L,\label{eq: Amplitude}
}where $e$ is the electric charge,  $\alpha_W\equiv \alpha /s_W^2$ with $\alpha$ the fine structure constant, and $s_W^2\equiv \sin^ 2 \theta_W$ the Weinberg angle. We also have introduced the notation
\eq{\Gamma_R\odot\Gamma^L\equiv [\bar{v}(p_1)P_R u(p_2)][\bar{u}(q_1)P_L v(q_2)],}
denoting a bi-spinor product, where $P_{L,R}=(1\mp\gamma_5)/2$ stand for the chirality projectors. The $F(s,t)$ function, in Eq. (\ref{eq: Amplitude}) is given as follows
\eq{F(s,t)&=\sum_{\alpha, \beta=1}^5\bigg[(B_{\mu\alpha}B_{\ell\beta}^*)^2\,(\mathcal{A}+\mathcal{A}')\label{eq: F definition}-2B_{\ell\alpha}^*B_{\mu\alpha}B_{\ell\beta}^*B_{\mu\beta}\,\left( \mathcal{B}+\mathcal{B}' \right)\bigg],
}where the factors $\mathcal{A}$, $\mathcal{A}'$, $\mathcal{B}$, and $\mathcal{B}'$ encode all the relevant one-loop tensor integrals given in terms of the Passarino-Veltman functions (see Appendix \ref{App: Loop functions}), they depend on two invariants, namely,  $s=(p_1+p_2)^2$ and $t=(p_1-q_2)^2$; and the masses of the internal particles involved in the loop computation. 

Taking the square of the amplitude in Eq. (\ref{eq: Amplitude}), it is straightforward to verify that the differential cross section for the $\mu^+(p_1)\mu^+(p_2)\to \ell^+(q_1)\ell^+ (q_2)$ $(\ell=e,\, \tau)$ transitions can be written as follows
\eq{
\frac{d\sigma}{dt}=\frac{1}{2}\,\frac{1}{64\pi}\left(\frac{\alpha_W}{16\pi}\frac{e^2}{s_W^2}\right)^2 \,|F(s,t)|^2\, ,\label{eq: Differential cross section}
}where the $1/2$ factor comes from the statistical property of having two indistinguishable particles in the final state and the invariant $t$ is evaluated in the range:
\eq{
t_{max(min)}=m_{\mu}^2+m_{\ell}^2-\frac{s}{2}\pm\frac{\lambda^{1/2}(s,m_{\mu}^2,m_{\mu}^2)\lambda^{1/2}(s,m_{\ell}^2,m_{\ell}^2)}{2 s}\, ,
}with $\lambda(x,y,z)=x^2+y^2+z^2-2(xy+xz+yz)$ the so-called K\"allen function \footnote{In the limit where the external particles are massless, we have $t\in(-s,0)$.}.

\section{Numerical Evaluation} \label{NE}

We now present an estimation for the $\mu^+\mu^+\to \ell^+\ell^+$ ($\ell=e,\, \tau$) cross section in the presence of heavy Majorana neutrinos. Our main goal is to establish the maximum cross-section for these processes caused by the presence of new heavy neutrino states in low-scale seesaw models. Therefore, we considered it appropriate to follow a conservative approach using the current global fit analysis of flavor and electroweak precision observables given in  \cite{Blennow:2023mqx, Celestino-Ramirez:2024gmq} \footnote{Notice that these limits depend not on the assumptions made on a particular model but solely on experimental data. Unlike the limits derived from cLFV processes like $\mu\to e\gamma$ or $\mu\to e$ conversion depend strongly on the masses considered for the new heavy neutrino states and on the specific textures of the neutrino mass matrix in different low-scale seesaw realizations.} 
\eq{
s_{\nu_e}<4.3\times 10^{-3},\quad s_{\nu_\mu}<1.61\times 10^{-2},\quad s_{\nu_\tau}<2.1\times 10^{-2}.\label{eq: Heavy-Light mixings limits}
}
Furthermore, we considered a perturbative limit assuming that the Yukawa couplings must satisfy the condition $|Y_{\nu i}|^2<4\pi$. In our framework, this translates into the relation
\eq{
r_{max}=\left(\frac{v\,  \sqrt{2 \pi}}{m_{N_1}\, \rm{max}\{s_{\nu_i}\}}\right)^{4}.\label{eq: r max}
}

\begin{figure}[!ht]
    \centering
    \begin{tabular}{cc}
    \includegraphics[scale=.44]{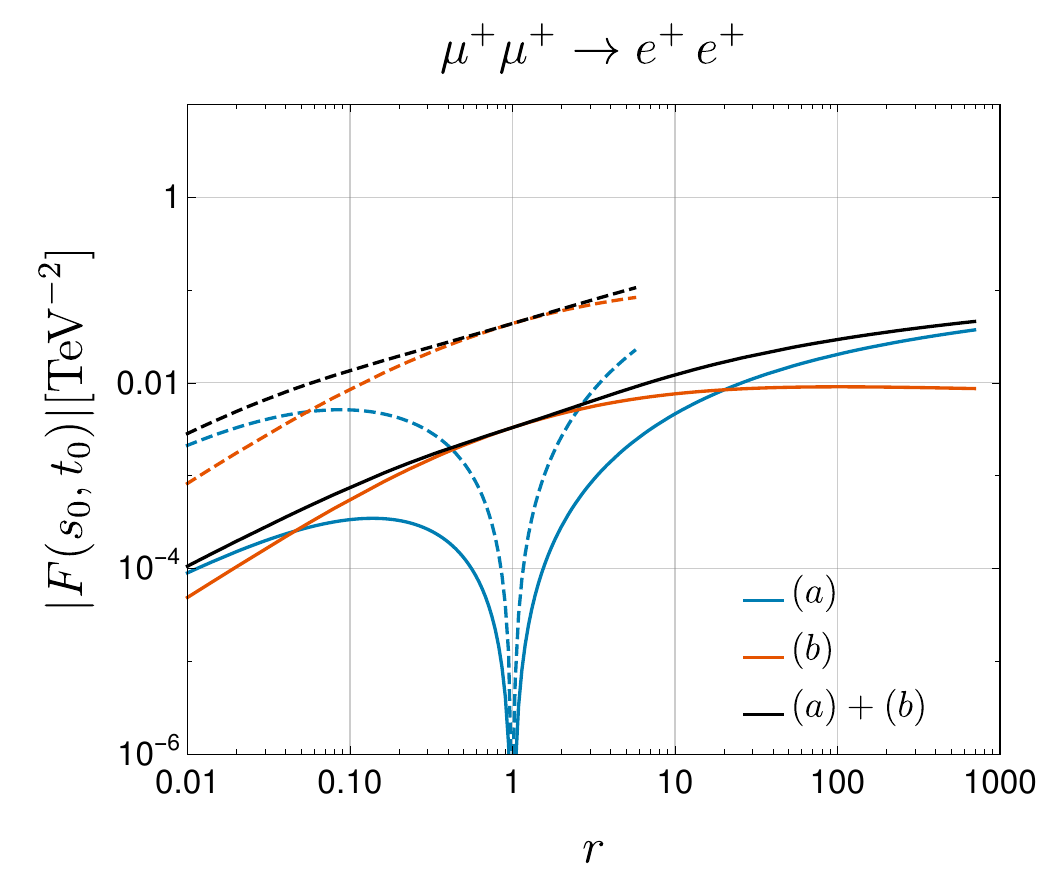} & \includegraphics[scale=.44]{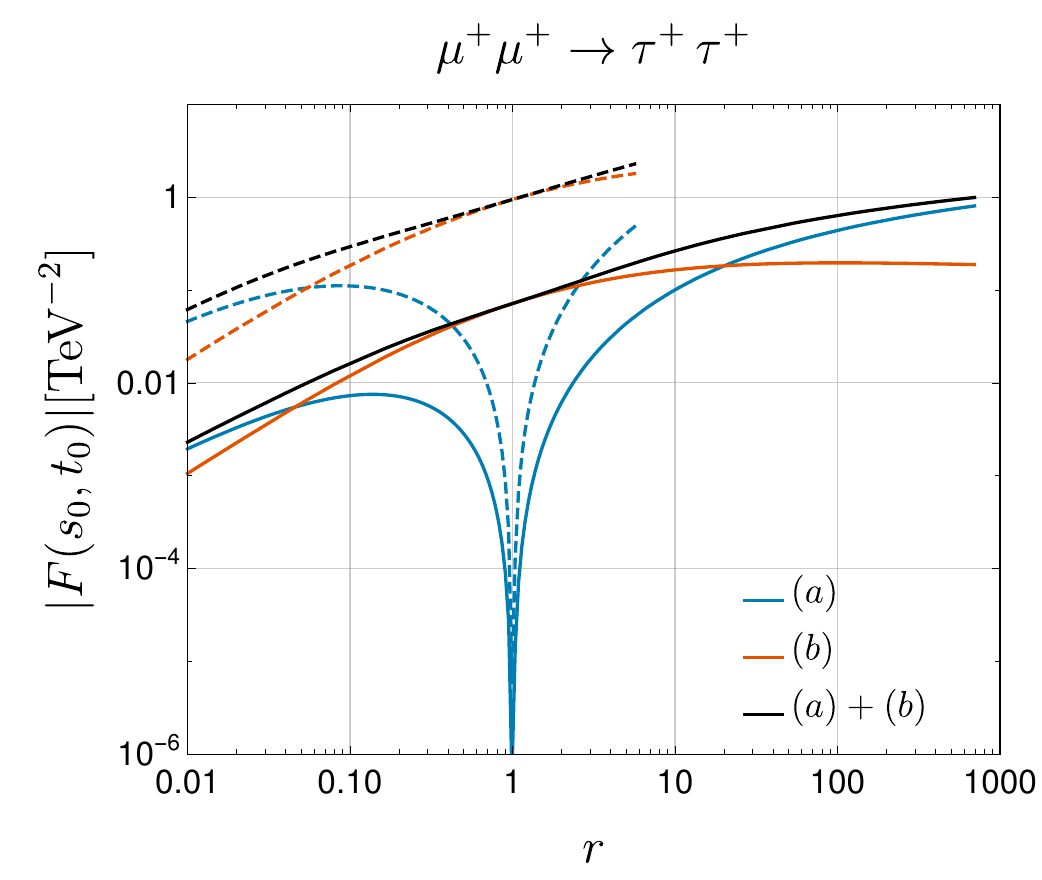} \\
       \quad\quad\quad\quad (a) &  \quad\quad\quad\quad (b)
    \end{tabular}
    \caption{$F(s,t)$ factor in terms of the mass splitting $r\equiv m_{N_2}^2/m_{N_1}^2$ of the two new heavy states. The heavy-light mixings are fixed by their maximum values according to expression (\ref{eq: Heavy-Light mixings limits}), and we also have considered that $(\sqrt{s_0},\sqrt{|t_0|})=(6, 0.1)$ TeV. The left (right) side of the figure illustrates the $\mu^+\mu^+\to e^+e^+$ $(\tau^+\tau^+)$ channel, where the behavior of the (a) and (b) contributions are represented by the cyan and orange lines, respectively; whereas the black line stands for the total contribution. The solid (dashed) line depicts a fixed value for $m_{N_1}=6$ ($20$) TeV. }
    \label{fig:F Factor}  
\end{figure}
Then considering Eqs. (\ref{eq: Heavy-Light mixings limits}) and (\ref{eq: r max}) we plot in Fig. \ref{fig:F Factor} the relevant $F(s,t)$ factor in terms of the mass splitting $r$ of the two new heavy states. Here, we used the maximum values from Eq. (\ref{eq: Heavy-Light mixings limits}) for the heavy-light mixings, along with a fixed point in the integration phase space set at $(\sqrt{s_0},\sqrt{|t_0|})=(6, 0.1)$ TeV. The left (right) side illustrates the $\mu^+\mu^+\to e^+e^+$ $(\tau^+\tau^+)$ channel, where the contributions of (a) and (b) are depicted by cyan and orange colors, respectively. Meanwhile, the black lines denote the total contribution, with solid (dashed) lines indicating the value for $m_{N_1}=6$ ($20$) TeV. Some noteworthy observations from this figure are as follows:

It turns out clear that in the limit when $r\to1$, the (a) contributions tend to zero, as expected, since in such a case the two new heavy states are degenerate forming a Dirac singlet field and recovering lepton number as a symmetry of our model. In such a case, the cross-section for the $\mu^+\mu^+ \to \ell^+\ell^+$ ($\ell=e,\,\tau$) transitions would be determined exclusively by the diagrams (b). This is a different result to previous estimations in  \cite{Cannoni:2002ny, Yang:2023ojm} where the contributions (b) were omitted. 

Note that for $m_{N_1}=6$ TeV and $r$ in the range $(0.1, 10)$, the  contributions (a) consistently remain below those from diagrams (b), and only for values of $r\gtrsim 11$ the contributions (a) become dominant.
Moreover, if we consider both the maximum values in (\ref{eq: Heavy-Light mixings limits}) for the heavy-light mixings and $m_{N_1}= 20$ TeV, then the upper limit of $r$ consistent with Eq. (\ref{eq: r max}) would be $r_{max}=4.64$ (this is depicted in Fig. \ref{fig:F Factor} where the dashed lines end).

Additionally, Fig. \ref{fig:F Factor}(b) illustrates a suppression of approximately two orders of magnitude for $\mu^+\mu^+\to e^+ e^+$  in comparison with the $\mu^+\mu^+\to\tau^+\tau^+$. This difference arises from the more restrictive experimental limits on the heavy-light mixings in Eq. (\ref{eq: Heavy-Light mixings limits}) involving electrons. Therefore, we will concentrate on the $\mu^+\mu^+\to\tau^+\tau^+$ channel. 

\begin{figure}[!ht]
    \centering
    \begin{tabular}{cc}
    \includegraphics[scale=.44]{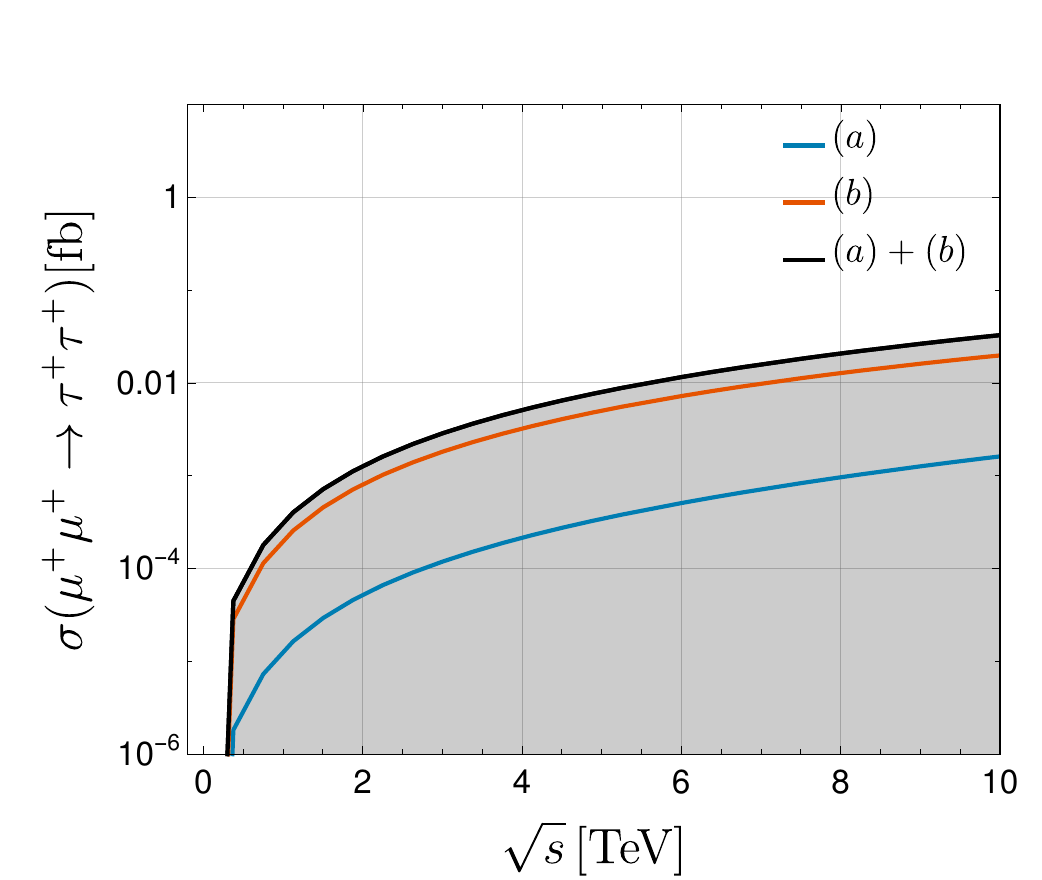} &     \includegraphics[scale=.44]{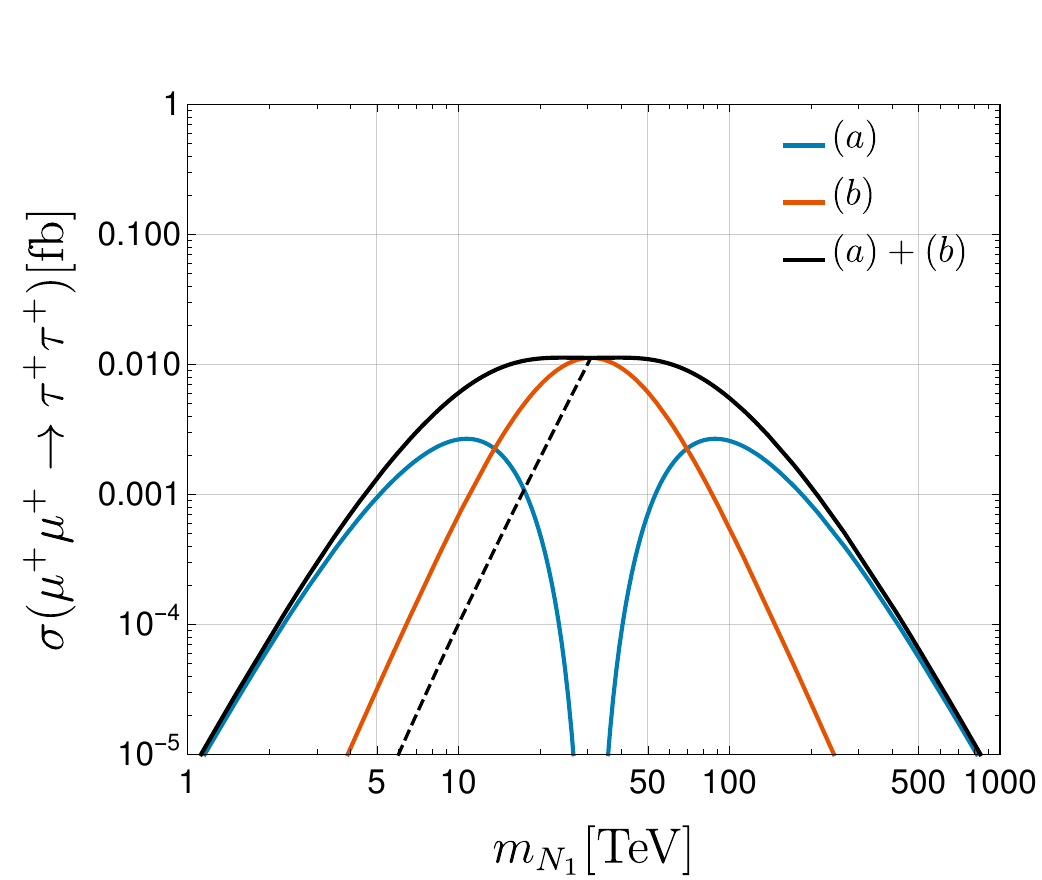} \\
       \quad\quad\quad\quad (a) &  \quad\quad\quad\quad (b)
    \end{tabular}
    \caption{Left (Right): Cross-section of the $\mu^+\mu^+\to\tau^+\tau^+$ transition as a function of $\sqrt{s}$  ($m_{N_1}$). The contributions of diagrams (a) and (b) are represented by the cyan and orange lines, respectively. The black lines stand for the total contribution. In the left plot, we have considered $m_{N_1}=20$ TeV and $r_{\text{max}}=4.64$ consistent with Eq. (\ref{eq: r max}). The shadowed area here represent the cross-section considering the heavy-light mixings below the current limits in expression (\ref{eq: Heavy-Light mixings limits}) and $r<r_{\textrm{max}}$.}
    \label{fig: Total cross section limits}  
\end{figure}

In Figure  \ref{fig: Total cross section limits}(a) we integrate Eq. (\ref{eq: Differential cross section}) over the $t$ invariant in order to plot the total cross section as a function of the invariant mass $\sqrt{s}$. Here, we have considered the mass of the heavy neutrino to be $m_{N_1}=20$ TeV and $r_{\text{max}}=4.64$ consistent with Eq. (\ref{eq: r max}). We observe from this that taking an energy of collision around $\sqrt{s}\approx6$ TeV, the maximal cross section for the $\mu^+\mu^+\to\tau^+\tau^+$ channel would be of the order $\mathcal{O}(10^{-2})$ fb. Therefore, if we consider the expected sensitivity reported in reference \cite{Hamada:2022uyn} for an expected total integrated luminosity (at $\sqrt{s}=6$ TeV) of around 12 fb$^{-1}$ year$^{-1}$, we would have approximately one $\mu^+\mu^+ \to\tau^+\tau^+$ event for every nine years of collisions. Nevertheless, following a more optimistic scenario, as presented in Ref. \cite{Fridell:2023gjx}, where the authors study similar transitions in the type-II seesaw model, namely, if we consider an integrated luminosity of $10$ ab$^{-1}$ with the center of mass energy $\sqrt{s}=10$ TeV, we estimated around $300$ events.

It is also important to stress that from a phenomenological point of view, these processes would be crucial to compare the genuine effects of heavy Majorana states with the contributions of new heavy Dirac singlets.
In this regard, Fig. \ref{fig: Total cross section limits}(b) depicts the behavior of the cross-section as a function of the heavy neutrino mass $m_{N_1}$ considering a transferred energy of $\sqrt{s}=6$ TeV, and  $r=r_{max}$ (solid black lines) \footnote{We considered $r=r_{max}$ to maximize the cross-section, taking $r<r_{max}$ would lead to lower values.}. The cyan and orange lines stand for the contributions of diagrams (a) and (b), respectively. Note that since the black line consistently lies above both the orange and cyan lines, the contributions of diagrams (a) and (b) in Figure \ref{fig: Diagrams} interfere constructively.

From Figure \ref{fig: Total cross section limits}(b), we can also see that for $m_{N_1}<13$ TeV and $m_{N_1}>65$ TeV, the dominant contribution arises from diagrams (a), which introduce explicit LNV vertices. Conversely, for $m_{N_1}$ within the interval $(13, 65)$ TeV, diagrams (b) become dominant, reaching a maximum of around $10^{-2}$ fb for $m_{N_1} \approx (20-50)$ TeV. Note that the cyan line tends to zero as $m_{N_1}$ approaches to 32 TeV because this corresponds to the scenario where $r_{\text{max}}=1$ and diagram (a) vanishes, as discussed previously. Finally, for comparative purposes, we include the Dirac singlet scenario ($r=1$), indicated by the black dashed line. In this case, $m_{N_1}<31$ TeV since beyond this range, $r=1$ contradicts eq. (\ref{eq: r max}). An important point to highlight here is that even in the absence of lepton number violation (LNV) in the theory, we still obtain non-zero values of the cross-section, reaching up to the limit of $\sigma\approx10^{-2}$ fb.

Notice that despite considering a simplified model, which, certainly, does not explain the masses and mixing of the light neutrino sector, this set-up provides both a clear analysis and a good estimation of the cLFV effects caused by the presence of new heavy neutrino states in low-scale seesaw scenarios. This point has been verified previously in references \cite{Hernandez-Tome:2019lkb, Hernandez-Tome:2020lmh}, where the authors study other cLFV transitions, such as $\ell\to \ell'\gamma$, $\ell\to \ell'\ell''\ell'''$, $\mu-e$ conversion in nuclei, $Z \to \ell\ell'$,  and $H \to \ell\ell'$ processes.

Let us recall that the main feature of low-scale seesaw models is that the Yukawa couplings of the new heavy neutrinos to the Standard Model (SM) leptons can be relatively large compared to those in conventional high-scale seesaw models. This is because the seesaw mechanism requires a balance between the scale of the new physics (the mass of the heavy neutrinos) and the Yukawa couplings to achieve the observed light neutrino masses. Moreover, in these low-scale scenarios, the physics responsible for neutrino masses could lie at the TeV scale, allowing significant mixing between the light neutrinos and the new heavy neutrinos, potentially leading to observable cLFV rates.

Because the contribution of the light-neutrino sector to cLFV is negligible, the relevant point to estimate cLFV effects in low-scale seesaw scenarios is to study the case of having large Yukawa couplings for the singlets. The simplicity of the set-up proposed in Refs. [75,76] let us understand this limit in a very simple way since all the phenomenology is described in terms of only $5$ parameters, namely, the three mixing angles $s_{\nu k}$ and the two $m_{N_1}$, and $m_{N_2}$ heavy neutrino masses (for which, simple analytical expressions have been derived).

\section{Conclusions}\label{Conclusions}

Some experimental collaborations expect to substantially improve the current limits on an extensive list of cLFV processes, including leptonic, hadron, and heavy boson decays \cite{MEGII:2018kmf, Cavoto, Blondel:2013ia, Alekou:2013eta, Kuno:2013mha}. These searches will either provide thrilling indications of new physics or impose even more stringent limits on extensions to the Standard Model.

In this work, motivated by the potential of future muon colliders, we investigate and revisit the computation of the  $\mu^+\mu^+\to \ell^+\ell^{+}$ $(\ell=\tau,\, e)$ processes in Type-I seesaw models, as part of complementary cLFV searches. We calculate the complete cross-section for these processes focusing on the so-called low scale scenarios. Our analysis is based on a setup involving only two new heavy Majorana neutrinos that allow us to study  the limit of pseudo-Dirac states, while taking into account the current constraints on their mixings with the light sector.

Our results show that the observation of the $\mu^+\mu^+\to \tau^+\tau^+$ cLFV transition due to the presence of new heavy Majorana neutrinos (with masses around $m_{N_1}\approx 20$ TeV) could be possible in a future same-sign muon collider, for an energy of collision $\sqrt{s} 	\gtrsim10$ TeV and an integrated luminosity of order $\mathcal{O}(\rm{ab}^{-1})$. 

The analysis presented in Figs. \ref{fig:F Factor} and \ref{fig: Total cross section limits} illustrate the relevance and behavior of contributions (a) and (b) to estimate the cross-section accurately in low-scale seesaw models. It turns out clear that for the contribution of pseudo-Dirac neutrinos, namely in the limit $(r\to 1$), contributions (b) dominate over contributions (a), because the latter tend to zero since lepton number is recovered as a symmetry. 
Nevertheless, notice that the maximum value for $r$ is determined for the perturbative Eq. (IV.2). Then, such as depicted in the plots, depending on the value fixed for $m_{N_1}$ you may have allowed values for $r$ where the contributions (a) become the dominant.  This distinction is crucial in distinguishing between the scenario with two nondegenerate heavy Majorana states and the degenerate case where they form a Dirac singlet field.

\section*{Acknowledgements}
D. P. S. thanks CONAHCYT for the financial support during his Ph.D studies. The work conducted by GHT received partial funding from CONAHCYT (SNII program). J.L.G.S. received  funding from the program estancias posdoctorales por M\'exico and from SNII program, both of CONAHCYT. J. R. acknowledges support from the program estancias posdoctorales por M\'exico of CONAHCYT and also from Universidad Autónoma de México-Programa de Apoyo a Proyectos de Investigación e Innovación Tecnológica Grant No. IG100322 and by CONAHCYT Grant No. CF-2023-G-433. 

\appendix

\section{Amplitudes}
\label{App: Amplitudes}

Before presenting the amplitudes of diagrams in Fig \ref{fig: Diagrams}, let us introduce the following notation to avoid cumbersome expressions:
\eq{
\Gamma_X^{\{1,\mu,\mu\nu\}}\odot\Gamma^Y_{\{1,\alpha,\alpha\beta\}}&\equiv \left[\bar{v}(p_1)\{1,\gamma^{\mu},\sigma^{\mu\nu}\}P_X u(p_2)\right]\,\left[\bar{u}(q_1)\{1,\gamma_{\alpha},\sigma_{\alpha\beta}\}P_Y v(q_2)\right],\\
\Gamma_X^{\{1,\mu,\mu\nu\}}\otimes\Gamma^Y_{\{1,\alpha,\alpha\beta\}}&\equiv \left[ \bar{u}(q_1)\{1,\gamma^{\mu},\sigma^{\mu\nu}\}P_X u(p_1)\right]\,\left[\bar{u}(q_2)\{1,\gamma_{\alpha},\sigma_{\alpha\beta}\}P_Y u(p_2)\right],
}where $P_X$ and $P_Y$ represent either the left (L) or right (R) projection operator.

\subsection{Diagrams with explicit LNV vertices}

The amplitudes for diagrams (a), namely, those with explicit LNV vertices can be expressed in a general form as follows
\eq{
\mathcal{M}_{LNV}^{a_1a_2}&=\frac{\alpha_W}{16\pi}\frac{e^2}{s_W^2} \sum_{\alpha, \beta}^5(B_{\mu\alpha}B_{\ell\beta}^*)^2\, M^{a_1a_2}_{LNV},
}where the superscript $a_1$ $(a_2)$ denotes the contribution of the particle ($W$ gauge boson or $\phi$ Goldstone boson) in the up (down) vertices of diagrams 1(a). We have found that  
\eq{
M_{LNV}^{WW}&=A_1\, \Gamma_R\odot \Gamma^L, \label{LNVww}\\
M_{LNV}^{\phi W+W\phi}&=A_2\, \Gamma_R\odot \Gamma^L + \Gamma_R^{\mu\alpha}\odot \Gamma^L_{\mu\beta} \left(B_{1}\,   q_{2_\alpha}p_1^{\beta}+B_{2}\,  q_{2_\alpha}p_2^{\beta} \right)\nonumber\\
&+\Gamma_R^{\mu\nu}\odot \Gamma^L\left( C_{1}\,p_{1_\mu}q_{2_\nu} +C_{2}\,p_{2_\mu}q_{2_\nu}\right)+ C_{3}\, \Gamma_R\odot \Gamma^L_{\mu\nu}\,p_{1}^\mu p_{2}^\nu,\label{LNVwppw}\\
M_{LNV}^{\phi \phi}&=A_3\, \Gamma_L\odot \Gamma^R.\label{LNVpp}
}
In the approximation limit where the masses of the external particles are identically zero, after some Dirac algebraic steps, Eq. (\ref{LNVwppw}) is simplified by using the following identities:
\eq{
\Gamma_R^{\mu\alpha}\odot\Gamma^L_{\mu\beta}\,q_{2_\alpha}p_{(1,2)}^\beta&=\pm q_2\cdot p_{(1,2)}\, \Gamma_R\odot\Gamma^L,\\
\Gamma_R^{\mu\nu}\odot\Gamma^L\, p_{(1,2)_\mu}q_{2_{\nu}}&=\mp i\, ( p_{(1,2)}\cdot q_2)\, \Gamma_R\odot\Gamma^L,\\
\Gamma_R\odot \Gamma_{\mu\nu}^L\,p_{1}^{\mu}\,p_{2}^{\nu} &=-i(q_1-q_2)\cdot p_2 \,\Gamma_R\odot\Gamma^L.
}
Therefore, the sum of expressions (\ref{LNVww}), (\ref{LNVwppw}), and (\ref{LNVpp}) is given by the simple expression
\eq{
M_{LNV}&=M_{LNV}^{WW}+M_{LNV}^{\phi W+W\phi}+M_{LNV}^{\phi\phi}, \nonumber\\
&=\mathcal{A}\, \Gamma_L\odot\Gamma^R\,,\label{LNV1t}
}where $\mathcal{A}=\left(A_1+A_{2T}+A_3\right)$, and we also have defined 
\eq{
A_{2T}&\equiv A_{2}+(q_2\cdot p_1)\left(B_{1}-i C_{1}\right)+(q_2\cdot p_2) \left(iC_{2}-B_2\right)-i(q_1-q_2)\cdot p_2\,C_3.
}

Similarly, the sum of the diagrams exchanging the final leptons ($\ell(q_1) \leftrightarrow  \ell (q_2))$ in Fig. (\ref{fig: Diagrams}) is identified straightforwardly from Eq. (\ref{LNV1t}) by the replacement $M_{LNV}\rightarrow M_{LNV}^{\prime}$, with
\eq{
M_{LNV}^{\prime}&=(-1)\mathcal{A}^{\prime}\, 
\Gamma_L\odot\Gamma^{\prime\,R}\,,\nonumber\\
&=(A_1^\prime+A_{2T}^\prime+A_3^\prime) 
\Gamma_L\odot\Gamma^R\label{LNV1pt},
} where $A'_i\equiv A_i\,(q_1\leftrightarrow q_2)$. The minus sign comes after considering the Feynman rules of Majorana fermions shown in Ref. \cite{Denner:1992vza}.  Therefore, all the contributions of diagrams (a) is given by 
\eq{
\mathcal{M}_{(a)}=\frac{\alpha_W}{16\pi}\frac{e^2}{s_W^2}   \sum_{\alpha, \beta}^5(B_{\mu\alpha}B_{\ell\beta}^*)^2 (\mathcal{A}+\mathcal{A'})\, \Gamma_R\odot\Gamma^L,\label{eq: Amplitude-a}
}

\subsection{Diagram (b) contributions}

The amplitudes of diagrams (b)  can be written in the generic form
\eq{
\mathcal{M}_{LNC}^{b_1b_2}&=\frac{\alpha_W}{16\pi}\frac{e^2}{s_W^2} \sum_{\alpha, \beta}B_{\ell\alpha}^*B_{\mu\alpha}B_{\ell\beta}^*B_{\mu\beta}\, M^{b_1 b_2}_{LNC},
}with 
\eq{
M_{LNC}^{WW}&=\alpha_{1}\,\Gamma^{\mu}_R\otimes\Gamma_{\mu}^R+\beta_{1}\,\Gamma^{\mu}_R\otimes \Gamma_{\nu}^R q_{2\mu}\,q_{1}^{\nu},\\
M_{LNC}^{\phi W+W \phi}&=\alpha_{2}\,\Gamma^{\mu}_R\otimes \Gamma_{\mu}^R,\\
M_{LNC}^{\phi\phi}&= \alpha_{3} \,\Gamma^{\mu}_R\otimes \Gamma_{\mu}^R+\left(\beta_{2}\,p_{2\mu}q_{1}^{\nu}+\beta_{3}\,q_{2\mu}\,q_{1}^{\nu}\right)\,\Gamma^{\mu}_R\otimes \Gamma_{\nu}^R.
}
Considering the approximation where the masses of the external particles are zero, the sum of the above expressions is given by 
\eq{
M_{LNC}&=M_{LNC}^{WW}+M_{LNC}^{\phi W+W\phi}+M_{LNC}^{\phi\phi}, \nonumber\\
&=\mathcal{B}\, \Gamma_R^\mu \otimes\Gamma^R_{\mu},\label{LNC1t}
}where, in the above expression, we have used the identity \footnote{Only valid in the massless limit.}
\eq{
\Gamma_R^\mu \otimes\Gamma^R_\nu\, q_{2\mu}q_{1}^{\nu}=\Gamma_R^\mu \otimes\Gamma^R_\nu\, p_{2\mu}q_{1}^{\nu}=(p_{2}\cdot q_{1})\Gamma_R^\mu \otimes\Gamma^R_\mu\,} and we have defined $\mathcal{B}\equiv\alpha_{1T}+\alpha_2+\alpha_{3T}$, where 
\eq{
\alpha_{1T}&=\alpha_{1}+(p_2\cdot q_1)\beta_1\\
\alpha_{3T}&=\alpha_3+(p_2\cdot q_1)(\beta_2+\beta_3).
}
Similar to the (a) contributions, the inclusion of the diagrams with the final charged leptons exchanged is added to the amplitude after the change $q_1\leftrightarrow q_2$ in the above expressions, which takes to
\eq{
\mathcal{M}_{(b)}=\frac{\alpha_W}{16\pi}\frac{e^2}{s_W^2} \sum_{\alpha, \beta}^5 B_{\ell\alpha}^*B_{\mu\alpha}B_{\ell\beta}^*B_{\mu\beta} (\mathcal{B}+\mathcal{B}')\, \Gamma_R^{\mu}\otimes\Gamma^R_{\mu}.\label{eq: Amplitude-b}
}

The sum of Eqs. (\ref{eq: Amplitude-a}) and (\ref{eq: Amplitude-b}) can be done immediately  by considering the Fierz identity
\eq{
\Gamma^{\mu}_R\otimes\Gamma_{\mu}^R=-2 \Gamma_R\odot\Gamma^L.
} Thus, the total contribution $\mathcal{M}=\mathcal{M}_{(a)}+\mathcal{M}_{(b)}$ is given by Eq. (\ref{eq: Amplitude}) .

\section{Loop functions}
\label{App: Loop functions}
All the relevant factors coming from the loop integration presented in the previous appendix are given in terms of Passarino-Veltman (PaVe) functions. We  have used the software Package X \cite{Patel:2015tea} for our computaion and we have found the following expression
\eq{
A_1&=4 m_{N_\alpha} m_{N_\beta}\overline{D}_0,\\
A_{2T}&=-2\frac{m_{N_\alpha}m_{N_\beta}}{m_W^2}\bigg[4\overline{D}_{00}-t\left(\overline{D}_0+\overline{D}_1+2 \overline{D}_2+\overline{D}_3\right.\label{eq: A2T}\left.+2(\overline{D}_{12}+\overline{D}_{13}+\overline{D}_{22}+\overline{D}_{23})\right)\bigg],\nonumber \\
A_3&=\frac{m_{N_\alpha}^3m_{N_\beta}^3}{m_W^4}\overline{D}_0.
}Note that these are the same expressions obtained in Ref.\cite{Cannoni:2002ny}, except for a minus sign in the eq.(\ref{eq: A2T}). We have introduced the following notation to express the argument of the Passarino-Veltman functions involved in the contributions of diagrams (a)
\eq{
\overline{D}_i\equiv D_{i}(0,0,0,0;s,t;m_W,m_{N_\alpha},m_W,m_{N_{\beta}}).
}
Regarding the diagrams (b) contributions we have found the relevant expressions
\eq{
\alpha_{1T}&= 4D_{00}-2t(D_0+D_1+D_3+D_{13})\,,\\
\alpha_2&=-2\frac{m_{N_{\alpha}}^2m_{N_{\beta}}^2}{m_W^2} D_0\,\\
\alpha_{3T}&= \frac{m_{N\alpha}^2m_{N\beta}^2}{2m_W^2}(2D_{00}-D_{13}),
}where, this time, the arguments of the Passarino-Veltman function are given by
\eq{
D_i\equiv D_i(0,0,0,0;u,t;m_{N_{\alpha}},m_W,m_{N_{\beta}}).
}
Defining the invariant $u\equiv(p_1-q_1)^2=(p_2-q_2)^2$ and taking $\mathcal{A}'$ ( $\mathcal{B}'$) factor related with $\mathcal{A}(\mathcal{B})$ by the change $t\leftrightarrow u$ in the corresponding PaVe functions.

\section{Expressions in terms of the massive states} \label{App: Heavy states}

Using the neutrino's mass framework introduced in section \ref{sec: the model}, we can rearrange the sum over all the mass eigenstates shown in eq.(\ref{eq: F definition}) in terms only of the heavy states as follows
\eq{
\sum_{\alpha, \beta=1}^5 B_{\ell\alpha}^*B_{\mu\alpha}B_{\ell\beta}^*B_{\mu\beta} d(m_{\alpha},m_{\beta})\equiv & \sum_{\alpha, \beta=4}^5 B_{\ell\alpha}^*B_{\mu\alpha}B_{\ell\beta}^*B_{\mu\beta}\left(d(m_{\alpha},m_{\beta})-d(0,m_{\beta}) \right.\nonumber\\
&\left. -d(m_{\alpha},0)+d(0,0)\right) \, \nonumber,
} where $d(m_{\alpha},m_{\beta})$ an arbitrary function that depends on the neutrino masses. It is important to remark that this simplification is valid only for this specific model where the light neutrinos have masses identically zero.

\bibliography{biblio}{}
\bibliographystyle{unsrt}

\end{document}